\documentclass[review,12pt]{elsarticle}

\usepackage{graphicx}
\usepackage{subfigure}
\usepackage{multirow}
\usepackage{wrapfig}
\usepackage{array}

\usepackage{tabularray}
\usepackage{tabularx}
\usepackage{rotating}
\usepackage{makecell}
\usepackage{booktabs}
\usepackage{csquotes}
\usepackage {subcaption} 
\usepackage{xcolor}

\usepackage[utf8]{inputenc}
\usepackage{amsmath}
\usepackage{amsthm}
\usepackage{amsfonts}
\usepackage{epsfig}
\usepackage{psfrag}
\usepackage{pstricks}
\usepackage{algorithm}
\usepackage{stfloats}

\usepackage{cuted}

\usepackage{algorithmicx}
\usepackage{algpseudocode}  
\floatname{algorithm}{Algorithm}

\graphicspath{{./Figures/}}

\usepackage{amssymb}
\usepackage{bm}
\usepackage{hyperref}
\hypersetup{
            colorlinks=true,
            linkcolor=blue,
            anchorcolor=blue,
            citecolor=blue
            }

\biboptions{comma,sort&compress}

\usepackage{enumitem}
\setlist{nosep,topsep=-\parskip}

\usepackage{ulem}

\journal{Elsevier}

\begin{document}

\begin{frontmatter}

\title{Semantic Direct Modeling}
\author[]{Qiang Zou\corref{cor}}\ead{qiangzou@cad.zju.edu.cn}
\author[]{Shuo Liu}

\cortext[cor]{Corresponding author.}
\address{State Key Laboratory of CAD$\&$CG, Zhejiang University, Hangzhou, 310058, China}

\begin{abstract}
Current direct modeling systems limit users to low-level interactions with vertices, edges, and faces, forcing designers to manage detailed geometric elements rather than focusing on high-level design intent. This paper introduces semantic direct modeling (SDM), a novel approach that lifts direct modeling from low-level geometric modifications to high-level semantic interactions. This is achieved by utilizing a large language model (LLM) fine-tuned with CAD-specific prompts, which can guide the LLM to reason through design intent and accurately interpret CAD commands, thereby allowing designers to express their intent using natural language. Additionally, SDM maps design intent to the corresponding geometric features in the CAD model through a new conditional, context-sensitive feature recognition method, which uses generative AI to dynamically assign feature labels based on design intent. Together, they enable a seamless flow from high-level design intent to low-level geometric modifications, bypassing tedious software interactions. The effectiveness of SDM has been validated through real mechanical design cases.
\end{abstract}




\end{frontmatter}


\section{Introduction}
Computer-aided design (CAD) has been widely adopted across industries such as automotive, aerospace, and biomedical engineering~\cite{li2020computer,zou2024meta}. One of its primary usages is the creation and modification of 3D computer models of products. The evolution of 3D CAD modeling can be broadly categorized into five stages: wireframe modeling~\cite{lee1999principles}, surface modeling~\cite{zhao2024tpms2step}, solid modeling~\cite{shapiro2002solid}, parametric modeling~\cite{shah1995parametric}, and direct modeling~\cite{zou2023variational}. As the latest paradigm in CAD, direct modeling enables history-free manipulation of CAD models, regardless of how the model was built~\cite{zou2019push}. Recognized as one of the most significant advancements in CAD over the past decade~\cite{ault2016direct}, direct modeling has been successfully integrated into leading commercial software such as Siemens NX, CATIA, PTC Creo, and Autodesk Inventor.

Despite improved flexibility and intuitiveness compared to parametric approaches, direct modeling still has room for improvement. In particular, current systems limit users to low-level interactions with CAD models, focusing primarily on vertices, edges, and faces within boundary representation (B-rep) models~\cite{zou2019push}. This way of working is not as attractive in the broader context of the design and manufacturing process~\cite{li2023xvoxel,zou2013iso}.  Designers, especially in the conceptual and process design stages, need to focus on high-level design semantics, such as slots or pattern features~\cite{shah1995parametric}. However, direct modeling often forces them to manage detailed geometric elements, rather than focusing on design intent. For example, editing a CAD model often involves tedious geometry selection and the precise specification and control of operations, which increases cognitive load and reduces efficiency. This highlights the need for semantic-level interaction in the human-direct modeling loop.

To address this, we introduce a semantic layer between the designer and the direct modeler, inspired by recent advancements in artificial intelligence (AI) and their applications to CAD~\cite{zhang2025diffusion,zou2025splinegen}. Specifically, we integrate speech recognition and large language models (LLMs) from the field of natural language processing (NLP) to create a speech-to-text module that robustly converts spoken commands into text. This enables designers to express their design intent through natural language (e.g., move the slot 3mm forward along with the X-axis), bypassing the need for tedious software interactions. The LLMs are used to analyze the text to extract semantic meanings, ensuring that the designer’s intent is well-captured and interpreted.

Building upon this natural expression of design intent, we further incorporate shape learning and feature recognition techniques to lift boundary representation (B-rep) models from purely geometric forms to semantic models (i.e., feature models). This enables the semantics extracted by the LLMs to be automatically mapped to the corresponding geometric portions of the CAD model being edited. Furthermore, LLMs can invoke appropriate direct modeling commands to modify the mapped geometry. In this manner, the human-to-semantics module and the geometry-to-semantics module converge, creating a seamless loop that directly links high-level design intent to low-level geometric modifications. We refer to this system as the Semantic Direct Modeling (SDM) System.

While SDM is easy to envision, its implementation presents unique challenges. Most notably, for a CAD model undergoing changes, a single face can simultaneously belong to multiple features, and the feature to which it is associated can vary depending on the design context~\cite{bidarra2000semantic}. This necessitates a dynamic, context-sensitive feature recognition method. While recent advances in deep learning have significantly enhanced feature recognition, e.g., UV-Net~\cite{jayaraman2021uv}, current methods typically assign static labels to features, without accounting for the design context~\cite{jayaraman2021uv,COLLIGAN2022103226,zhang2024brepmfr,wu2024aagnet,lambourne2021brepnet}. This limitation arises because these methods are largely adapted from semantic segmentation techniques in computer vision, which rely on discriminative AI models that are inherently static. As such, these methods are not well-suited to the problem in this paper. To address this, we propose a conditional feature recognition method that leverages generative AI to dynamically generate feature labels based on the semantics provided by LLMs.

The main contributions of this paper are as follows:
\begin{itemize}
    \item A semantic direct modeling framework that lifts direct modeling from low-level geometric modifications to high-level semantic interactions, enabling a seamless flow from designer intent to CAD modeling.
    \item A dynamic, context-sensitive feature recognition method that uses generative AI to dynamically generate feature labels based on design intent.
    \item A CAD domain-specific prompt engineering method that guides LLMs to reason through design intent and interpret CAD commands, mimicking the designer's thought process.
\end{itemize}

The remainder of this paper is organized as follows. Section.~\ref{sec:related_work} reviews the relevant literature, Section.~\ref{sec:methods} elaborates on the methods of SDM, and Section.~\ref{sec:results} presents validation of the method through a series of examples and comparisons. Conclusions are provided in Section.~\ref{sec:conclusion}.

\section{Related Work}
\label{sec:related_work}
\subsection{Direct modeling}
Direct modeling is a 3D CAD technique that allows users to directly and freely manipulate the geometric elements (e.g., boundary faces) in a B-rep model, which emerged around 2010. This method offers three major benefits: flexible edits, fast updates, and intuitive interaction~\cite{zou2023variational}. It is considered one of the recent major advancements in CAD modeling, following the parametric CAD paradigm~\cite{zou2019push}.

The origins of direct modeling can be traced back to the 1970s when it was known as tweaking, a specific type of local operations~\cite{grayer1980alternative,stroud2006boundary}. At that time, however, tweaking was restricted by its inability to violate the B-rep topology. In contrast, modern direct modeling removes this limitation, enabling greater flexibility in CAD modeling~\cite{fahlbusch1995hp}. While this flexibility offers advantages, it also introduces challenges: the potential disruption of topology and the breaking of parametric relationships in CAD models that were intact before direct edits~\cite{rossignac1990issues}. Significant efforts from both academia and industry have sought to address these challenges, with promising solutions demonstrated in academic research~\cite{zou2019push,zou2022robust,lipp2014pushpull,zou2020decision,pan2022automatic,zou2019variational} and in industry-leading software like Ansys SpaceClaim and Siemens NX.

While much attention has focused on topology and parametric disruption, another critical issue remains underexplored. Direct modeling limits users to low-level interactions with CAD models, focusing on vertices, edges, and faces within B-rep models~\cite{ault2016direct,zou2019push}, and consequently, the cognitive load increased and design efficiency reduced. This limitation is particularly relevant in the emerging era of intelligent CAD~\cite{zou2024intelligent}. This gives rise to the need for semantic-level direct modeling, which enhances interactions with higher-level semantic information and improves efficiency. This is the focus of this paper.

\subsection{Deep learning in CAD}
Building on the success of deep learning in NLP and CV, researchers, particularly in the graphics community, have explored its potential for processing 3D models. These research studies have inspired our use of LLMs and generative AI to develop the SDM system. For this reason, a brief review of CAD-relevant works is provided below.

\textbf{CAD-oriented Discriminative Networks.} Discriminative deep learning methods are primarily applied to feature recognition tasks in CAD. Classical networks, such as convolutional neural networks (CNNs), graph neural networks (GNNs), and Transformers, have all been employed for this purpose. For example, UV-Net~\cite{jayaraman2021uv} and Hierarchical CADNet~\cite{COLLIGAN2022103226} use CNNs to learn geometric features and GNNs to model topological patterns, achieving high feature recognition accuracy. Building on these early works, BRepGAT~\cite{lee2023brepgat} leverages GAT (Graph Attention Network) to more effectively capture topological information compared to standard GNNs. BRepNet~\cite{lambourne2021brepnet} and BrepMFR~\cite{zhang2024brepmfr} enhance CNNs with more semantic geometric features, such as surface types and edge convexity, resulting in improved classification performance. BRep-BERT~\cite{Lou2023brepbert} introduces the concept of BERT into B-rep learning, demonstrating superior performance in few-shot learning. AAGNet~\cite{wu2024aagnet} integrates the attention mechanism from the Transformer architecture with GNNs to effectively recognize machining features. While these networks are effective in many applications, they typically assign static labels to features, without considering the dynamic design context. As a result, their application to the interactive SDM problem is limited.
 
\textbf{CAD-oriented Generative Networks.}
While there is extensive research dedicated to applying generative AI to geometric modeling, it is primarily used to generate graphical models such as point clouds and meshes~\cite{liu2023meshdiffusion, chen2025meshxl,luo2021diffusion,shu20193d}. Only a few studies focus on CAD models, particularly B-rep models. Typical methods include BrepGen~\cite{xu2024brepgen}, ComplexGen~\cite{guo2022complexgen}, SolidGen~\cite{jayaraman2023solidgen}, and DeepCAD~\cite{wu2021deepcad}. These methods generally focus on generating geometric parameters and topological specifications of B-rep models, or the construction history (typically sketch-extrusion sequences) of B-rep models from various inputs like partial models or text prompts. For example, BrepGen~\cite{xu2024brepgen} uses diffusion models (a type of generative model) to generate B-rep models from sampling noise. To build SDM, we seek a generative method that can dynamically generate feature labels based on varying design contexts. Therefore, the current CAD-oriented generative methods are not suitable, and specialized development is needed.

\begin{figure*}[b]
    \centering
    \includegraphics[width=0.95\linewidth]{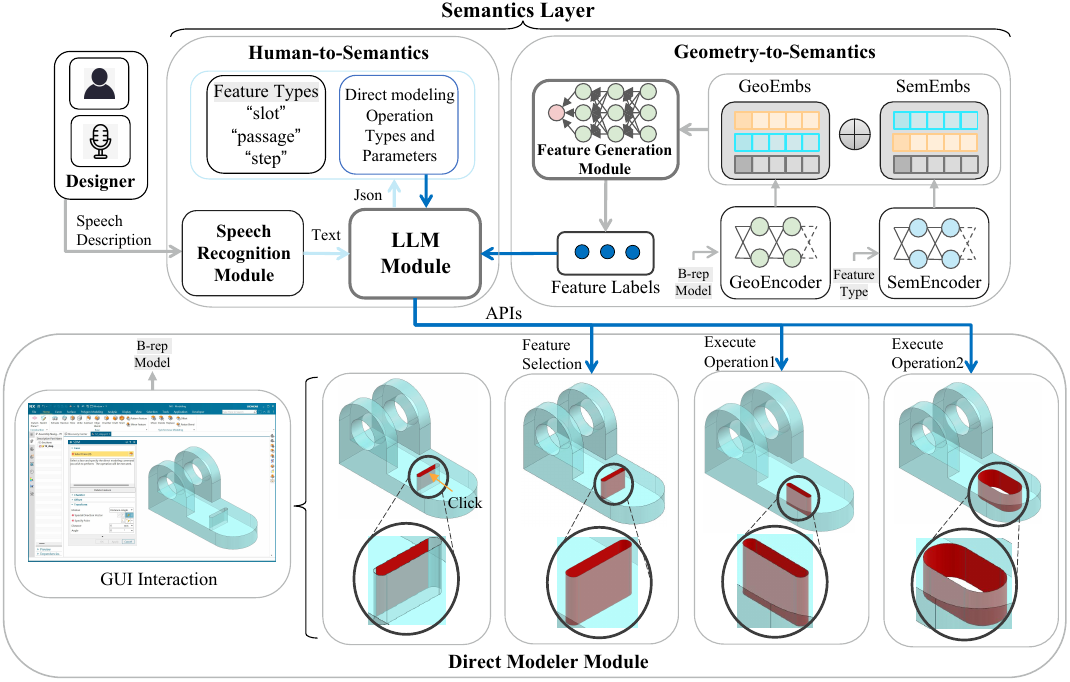}
    \caption{The overall architecture of SDM.}  
    \label{fig:overall}
\end{figure*}

\section{Methods}
\label{sec:methods}
To lift direct modeling from low-level geometric modifications to high-level semantic interactions, an additional semantic layer is introduced between the designer and the direct modeler. This layer interfaces with the designer through speech recognition and large language models (LLMs), enabling designers to express their design intent in natural language. It connects to the direct modeler by automatically generating relevant API calls  using LLMs. Section~\ref{sec:overall} outlines the architectural design, with further details on the constituent modules provided in Sections~\ref{sec:Prompt-engineering}, \ref{sec:feature-generation}, and~\ref{sec:training}.

\subsection{The overall system design}
\label{sec:overall}
Designing SDM requires careful consideration of the inherent differences between designers, LLMs, and direct modelers. Designers communicate design commands through speech, while LLMs process only text. Additionally, the text generated by LLMs is not directly understandable by direct modelers, and its relevance to the B-rep model is often unclear. To bridge these gaps, the system must translate the designer’s speech input into a format that the direct modeler can understand and act upon. This requires developing a system that can map the designer’s intent into precise modeling operations, ensuring seamless communication between all components.

Bridging the gap between the designer and the LLMs is straightforward with the addition of speech recognition. However, connecting the LLMs to the direct modeler is more complex. We need: (1) an interpreter to extract CAD semantics from speech-converted text, (2) a dynamic feature generation module to help the LLMs correctly identify the region of interest for model modifications, and (3) an API generation module to produce proper calls for making model edits aligned with the designer's intent. The overall system architecture for integrating these modules is shown in Fig.~\ref{fig:overall}. The details of each component are outlined below.

\textbf{Direct Modeler Module.} The direct modeler is responsible for executing any model edits implied by the designer's speech commands, potentially combined with GUI interactions. Siemens NX~\cite{nx} was chosen for this task after evaluating several leading direct modelers, including ANSYS SpaceClaim~\cite{ansys}, PTC Creo~\cite{ptccreo}, and Autodesk Inventor~\cite{autodesk}. NX was found to offer the most robust performance~\cite{zou2022robust}, which is crucial given the potential vagueness of speech commands. Additionally, NX provides comprehensive developer documentation for plugin extension and customization.

\textbf{LLM Module.} The task here is to interpret speech-converted text and extract its corresponding CAD semantics, organizing the information into a structured format. This involves identifying operation types, parameters, and the features upon which the operations are applied, followed by generating appropriate API calls. Since speech-converted text is often vague and error-prone, the LLMs must be robust enough to handle inconsistencies. It should also be capable of learning from developer documentation to generate accurate API calls. Most importantly, the LLM system should be open-source. Given these criteria, we have chosen Llama~\cite{dubey2024llama} for its excellent performance and open-source nature. 

However, it is important to note that while Llama performs well for general applications, it lacks domain-specific knowledge of CAD modeling, which can lead to suboptimal performance when interpreting CAD-related commands. We provide a solution to this unique challenge by designing CAD-specific prompts and teaching the LLM to break down tasks into smaller, more manageable reasoning steps that LLMs can process with only general knowledge. Details of this solution are provided in Section~\ref{sec:Prompt-engineering}.

\textbf{Speech Recognition Module.} Speech recognition is a well-established field, with systems such as Whisper~\cite{radford2023robust} demonstrating high accuracy and efficiency. For instance, Whisper has achieved a word error rate as low as 3\%, making it highly reliable for developing the SDM system. In this work, Whisper is the system of choice.

\textbf{Dynamic Feature Generation Module.} This module uses the semantics (primarily feature types) extracted by the LLM as conditions to identify the relevant boundary faces within the B-rep model, which are then prepared for use in subsequent model edits. The key challenge lies in the fact that semantics are textual, while boundary faces are geometric, two fundamentally different modalities. We provide a novel solution to this challenge by using the cross-attention mechanism~\cite{vaswani2017attention} to align and process textual and geometric data simultaneously. The combined multimodal embeddings are then fed into the Transformer decoder to generate the IDs of the relevant boundary faces. Details of this method can be found in Section~\ref{sec:feature-generation}.

\textbf{API Generation Module.} This task is also handled by the LLM module. It takes as input the structured representation of semantics (from the LLM) and the IDs of the relevant boundary faces (from the dynamic feature generation module), and then transforms them into the appropriate format for API calls. These calls then drive the direct modeler to execute the final model edits. While this task may seem complex at first, it is relatively straightforward due to the current ability of LLMs to generate code effectively~\cite{jiang2024survey}, which closely mirrors the API generation task in this work. Additionally, NX offers extensive developer documentation, which LLMs can directly process.

\subsection{CAD domain-specific prompt engineering} 
\label{sec:Prompt-engineering}
\begin{figure}[b]
    \centering
    \subfigure[Unprompted]{
        \includegraphics[width=0.4\textwidth]{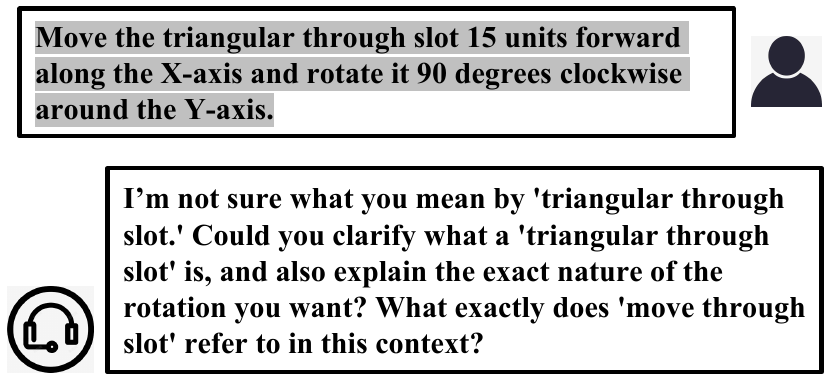}  
        \label{fig:unprompted}
    }
    \subfigure[Simple prompt]{
        \includegraphics[width=0.4\textwidth]{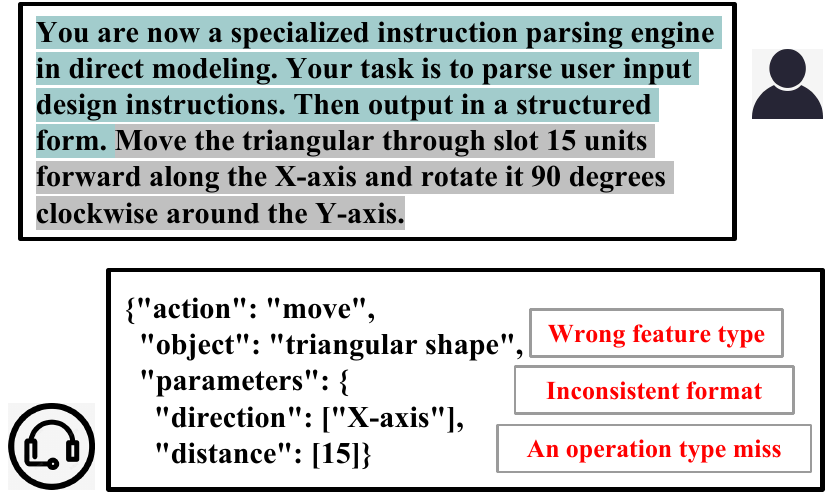}   
        \label{fig:simple-prompt}
    }
    \caption{Comparison of unprompted and simple prompt responses for CAD instruction comprehension.}
   
\end{figure}

LLMs are typically trained on general-purpose datasets, which provide broad knowledge but lack specialization in CAD-specific tasks~\cite{raiaan2024review}. As a result, directly applying LLMs to CAD modeling commands often leads to unsatisfactory outcomes. As can be seen from the example in Fig.~\ref{fig:unprompted}, the raw Llama model struggles to understand CAD terminologies, let alone interpret them into a structured representation of semantics. While prompts~\cite{schick2020exploiting,brown2020language} can improve results by offering contextual guidance, such as specifying the task type or output format, they cannot fully address the problem. As illustrated in Fig.~\ref{fig:simple-prompt}, simple prompts may help, but they fail to equip the LLMs with the precision and robustness needed to comprehend complex CAD workflows and implicit design intent. This highlights the need for more sophisticated prompts to adapt LLMs to the CAD domain.

To guide the LLMs in producing desired outputs, we employ the Chain-of-Thought (CoT) method~\cite{wei2022chain}, which enables the model to break down complex tasks into smaller, manageable components by following intermediate reasoning steps. This approach mimics the human thought process when navigating through complex problems. By adapting this approach to SDM, we enable the LLMs to reason through design intent step-by-step and generate more accurate, consistent, and structured outputs.

\begin{figure}[b]
    \centering
    \includegraphics[width=1\linewidth]{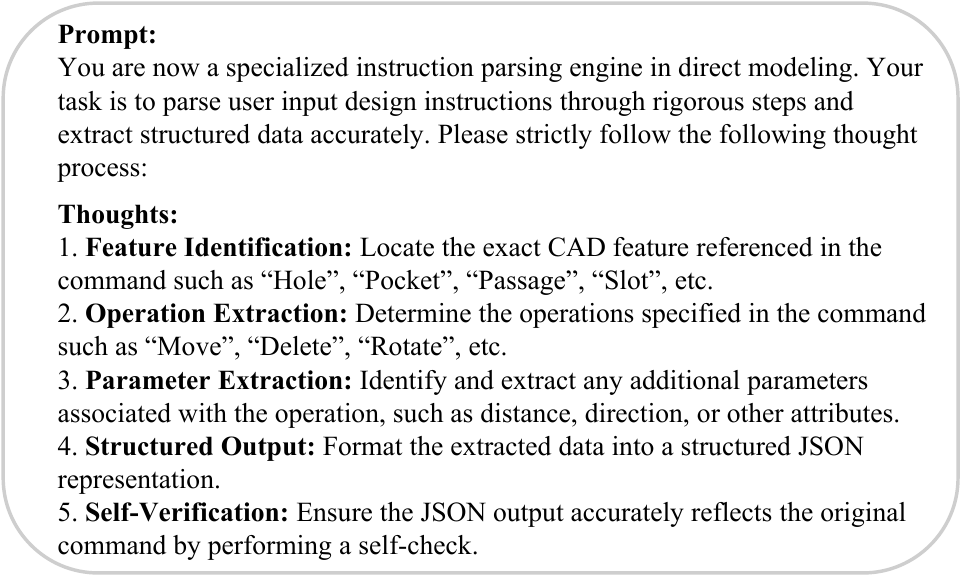}
 \caption{Chain-of-Thought guided structured parsing process for CAD design instructions.}
    \label{fig:CoT}
\end{figure}

As illustrated in Fig.~\ref{fig:CoT}, thought processes based on the framework of CoT have been designed. These processes guide the LLMs to autonomously decompose complex designer instructions through five key steps, thereby progressively emulating the reasoning logic of human designers. Given that the operations executed by designers are based on specific geometric entities, the first step extracts the target CAD feature type implied in the command. Subsequently, the operation types and parameters are sequentially extracted. To augment the model's comprehension, CAD-specific terminology is integrated into the first three steps. With this information, the model then formulates the output in a structured format, opting for JSON due to its unambiguous structure and ease of post-processing. As the final step, the model conducts self-verification to ensure it accurately mirrors the original command's intent.

\begin{figure}[t]
    \centering
    \includegraphics[width=1\linewidth]{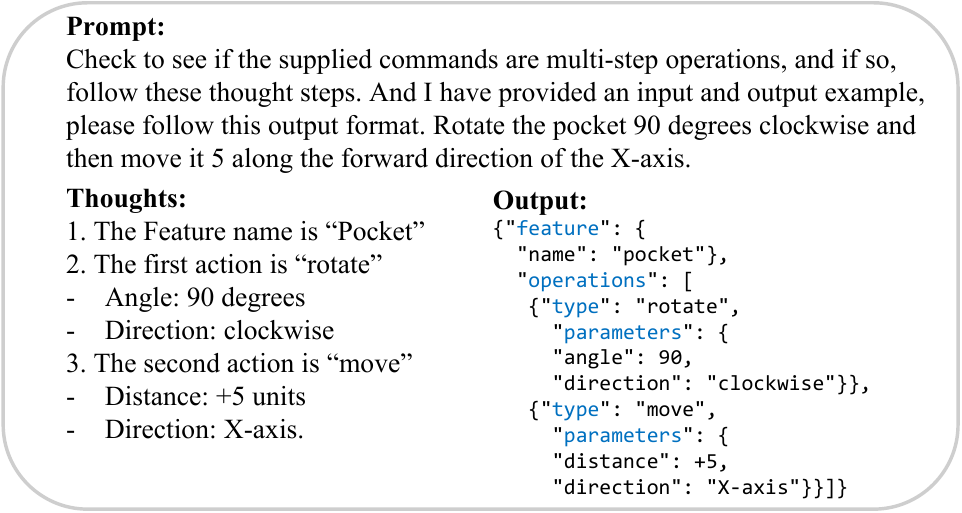}
    \caption{Multi-step instruction parsing and providing the few-shots to enhance the ability of LLM in CAD domain tasks.}
    \label{fig:Few-Shots}
\end{figure}

A single design command often encompasses multiple operations, necessitating the generation of structured outputs that precisely reflect each step. To achieve this, we introduce a multi-step reasoning template, as illustrated in Fig.~\ref{fig:Few-Shots}. Additionally, few-shot examples are provided to guide the model in generating standardized structured outputs, effectively mapping its thought processes to well-defined output formats. This approach ensures consistency and enhances usability in downstream CAD applications.

\subsection{Dynamic feature generation}
\label{sec:feature-generation}
The task here is to perform text-conditioned feature generation, specifically in the form of relevant boundary face IDs. To achieve this, we first learn vector representations of both texts and B-rep models and then align them using a cross-attention mechanism. The resulting fused embeddings are subsequently passed through the Transformer decoder to generate the IDs of the relevant boundary faces.

\subsubsection{Textual embedding for design semantics}
\label{sec:textual-embedding}
After the LLM processes speech-converted text, the design semantics implied in this text are transformed into a structured format comprising operation types, parameters, and feature types. Feature types will serve as conditions to guide the generation of the correct boundary face IDs for the intended feature. (Note that multiple features may share the same type in a B-rep model. To resolve this ambiguity, additional information is required, most importantly the spatial location of the feature. This information is obtained by asking the designer to perform a simple mouse click to indicate where the feature is located.)

Feature types, represented in textual form, are not directly interpretable by generative models. To enable processing by these models, feature types must first be transformed into vector representations (also referred to as embeddings), where similar features are mapped to nearby vectors in the embedding space. This enables the model to recognize and differentiate between different features, facilitating the correct generation of boundary face IDs. The specific embedding method employed in this work is the text-embedding-3-small from OpenAI. It is a pre-trained model that maps text (e.g., ``Rectangular Through Slot") into a 256-dimensional embedding $\mathbf{E}_\mathbf{S}$.

\subsubsection{Geometric embedding for B-rep models}
\label{sec:geometric-embedding}
Inspired by the success of Transformers in NLP and CV, where they significantly outperform traditional networks like CNNs and GNNs, we also use Transformers to carry out geometric embedding for B-rep models. To the best of our knowledge, no prior research has explored this approach, so we present one below.

To enable Transformers to process B-rep models, we must first convert the model into a sequence of tokens that can be effectively interpreted by these models. A token represents a small, meaningful unit of data that AI models can process~\cite{dosovitskiy2020image}. In the context of B-rep models, tokens correspond to the vector representations of boundary elements of vertices, edges, and faces. Considering that there is an inherent hierarchy among vertices, edges, and faces in B-rep models---vertices are connected by edges, which in turn define the boundaries of faces---the tokens representing these elements should not be treated as independent, but instead organized in a structured manner to preserve their relationships.

\begin{figure}[t]
    \centering
    \includegraphics[width=0.95\linewidth]{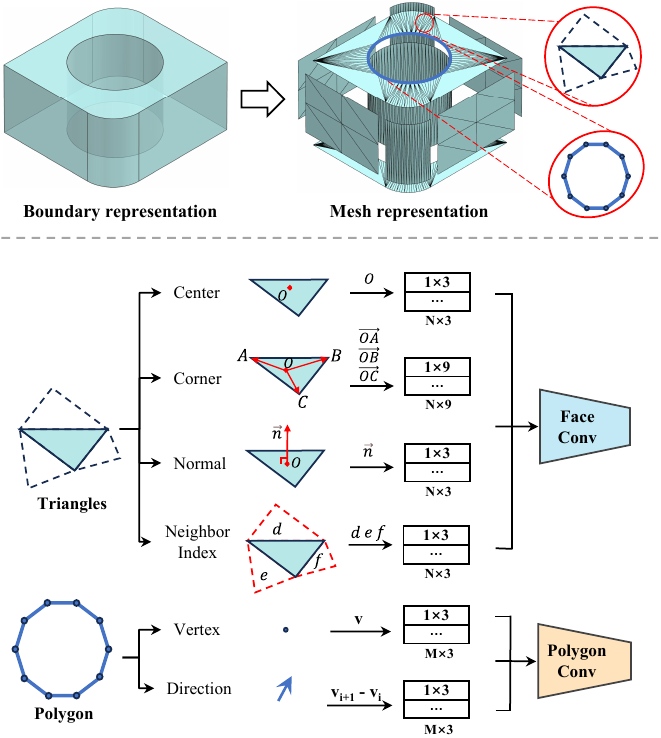}
    \caption{Illustration of B-rep model tokenization.}
    \label{fig:brep_process}
\end{figure}

To achieve this, our approach begins by converting a B-rep model into a triangular mesh, resulting in a collection of triangles for each face of the B-rep model, multiple line segments for each edge, and a polygon for each edge loop, as illustrated in Fig.~\ref{fig:brep_process}. Next, we generate tokens for each edge, loop, and face, and aggregate them into compound, face-wise tokens according to the B-rep model's hierarchical topology. These tokens are then input into the Transformer encoder to give the geometric embedding of the overall B-rep model. Note that tokenizing vertices is not necessary because they can be implied by the tokens of edges.

For a segment $\mathbf{s}=\{\mathbf{v}_1,\mathbf{v}_2\}$, where $\mathbf{v}_1$ and $\mathbf{v}_2$ are the start and end vertices, respectively, we construct its token by concatenating its start vertex's coordinates with its direction $\mathbf{d} = \mathbf{v}_{2} - \mathbf{v}_{1}$:
\begin{equation}
    \mathbf{Token}(\mathbf{s})=[\mathbf{v}_1;\mathbf{d}_1]\in\mathbb{R}^{6}.
\end{equation}

For a polygon $\mathbf{p}=\{\mathbf{v}_1,\mathbf{v}_2,\ldots,\mathbf{v}_n\}$, we construct its token by aggregating segment tokens, as follows:
\begin{equation}
    \mathbf{Token}(\mathbf{p})=[\mathbf{v}_1;\mathbf{d}_1;\mathbf{v}_2;\mathbf{d}_2;\ldots;\mathbf{v}_n;\mathbf{d}_n]\in\mathbb{R}^{n\times6}.
\end{equation}

For a face $\mathbf{f}$ consisting of a collection of triangles $\{\mathbf{t}_1,\mathbf{t}_2,\ldots,\mathbf{t}_n\}$, we first tokenize each triangle and then aggregate them into the face token by concatenating these triangle tokens, adapted from the method presented in~\cite{feng2019meshnet}. Specifically, a triangle's token comprises four attributes:
\begin{itemize}
    \item \textbf{Center} $( C \in \mathbb{R}^3)$: The center coordinates of the triangle.
    \item \textbf{Corner} $(D_1, D_2, D_3 \in \mathbb{R}^3)$: Direction vectors from the center to each vertex, as shown in Fig.~\ref{fig:brep_process}.
    \item \textbf{Normal} $(N \in \mathbb{R}^3)$: The unit normal vector of the triangle.
    \item \textbf{Neighbor Index} $(NI \in \mathbb{Z}^3)$: Indices of the connected triangles (the triangle is filled with the index of itself if it is connected to fewer than three adjacent triangles).
\end{itemize}
Because the Center attribute describes the triangle's global location, while the others describe its local shape, we use two tokens to represent a single triangle:
\begin{equation}
\begin{aligned}
\mathbf{Token}(\mathbf{t})^{ {location}} &=\left[ {C}\right] \in \mathbb{R}^{1 \times 3},\\
\mathbf{Token}(\mathbf{t})^{ {shape}} &=\left[ {N}; {D}_1;{D}_2; {D}_3; {NI}\right] \in \mathbb{R}^{1 \times 15}.
\end{aligned}
\end{equation}

The overall face token can then be obtained by simply arranging the triangle tokens into an array. However, this approach only provides low-level geometric coordinates to the Transformer, without capturing any high-level geometric features of the face. To address this, we apply convolution operations to the array of triangle tokens to extract higher-level features that better represent the face's characteristics. These convolution operations are not new but referenced from existing work, so we omit the details here. For more information, please refer to MeshNet~\cite{feng2019meshnet}. Similarly, the tokens of the previous polygons undergo a similar convolution process.

\begin{figure*}[t]
    \centering
    \includegraphics[width=0.98\linewidth]{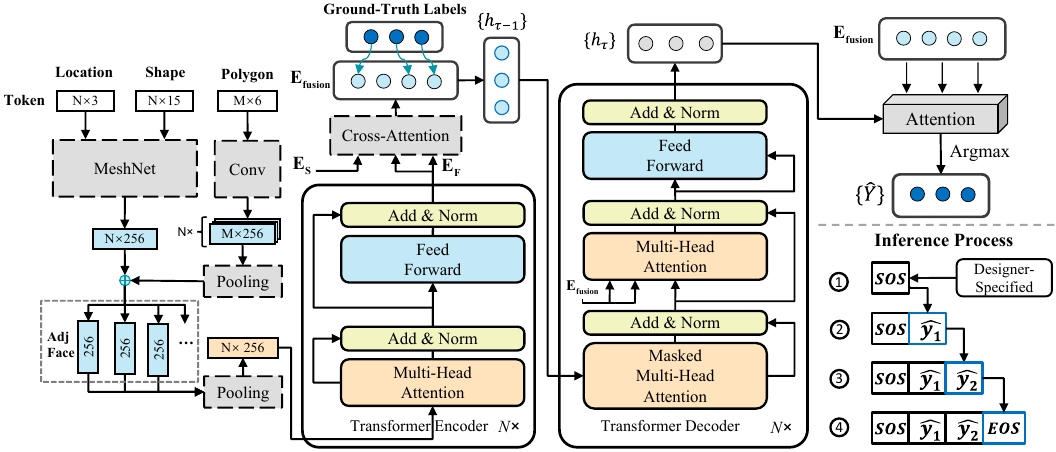}
    \caption{Illustration of feature generation module.}
    \label{fig:Network}
\end{figure*}

Next, we hierarchically aggregate tokens of individual polygons and faces (Fig.~\ref{fig:Network}). Since a face may have multiple boundary loops, i.e., polygons in the mesh representation, we first apply a sum pooling operation to the tokens of these boundary polygons to obtain a combined token. This token is then added to the face's token to capture their topological relationship. 

In addition to the above aggregation, we also apply a sum pooling operation to the tokens of neighboring faces. We do this because features (e.g., a hole or slot) are often composed of a group of adjacent faces. By allowing a face's token to be aware of its neighboring faces' tokens, we facilitate the Transformer encoder in learning not only the feature's local geometric characteristics but also the topological relationships among the boundary faces belonging to the feature. We then obtain the embedding of each face of the B-rep model, denoted as $\mathbf{E}_\mathbf{f}$. The embeddings of all faces, which also serve as geometric embeddings, are denoted as $\mathbf{E}_\mathbf{F}$.

\subsubsection{Textual-geometric embedding fusion}

Having obtained textual embeddings $\mathbf{E}_\mathbf{S}$ of the design semantics and the geometric embeddings $\mathbf{E}_\mathbf{F}$ of B-rep models, we use the cross-attention mechanism~\cite{vaswani2017attention} to align these data. The cross-attention process can be formulated as: 
\begin{equation}
\begin{aligned}
        \mathbf{E}_{\mathrm{fusion}} &= \text{Cross-Attention}(\mathbf{E}_\mathbf{S},\mathbf{E}_\mathbf{F})\\&=
    \mathrm{softmax}\left(\frac{(\mathbf{E}_\mathbf{S}W_Q)(\mathbf{E}_\mathbf{F}W_K)^T}{\sqrt{D_k}}\right)(\mathbf{E}_\mathbf{F}W_V).
\end{aligned}
\end{equation}
Here, $\mathbf{E}_\mathbf{S}$ serves as the Query, while $\mathbf{E}_\mathbf{F}$ providing the Key and Value. The resulting aligned representation $\mathbf{E}_{\mathrm{fusion}}$ captures the relationships between the textual semantics and geometric features, enabling the model to effectively fuse information from both modalities. 

\subsubsection{Conditional feature generation}
 
In our system, this module generates the feature label sequence conditioned on the semantic and geometric data provided by the designer. As previously mentioned, given the superiority of the Transformer architecture, we leverage the Transformer decoder as our decoder backbone. The input to the transformer decoder consists of two components: (1) the textual-geometric embedding fusion embeddings $\mathbf{E}_{\mathrm{fusion}}$, which encode the geometric and textual features of the input data, and (2) the sequence of embeddings representing the partially generated face IDs up to the current time step. 

Since the number of faces composing each feature varies, a key challenge lies in ensuring the number of generated faces. To address this, the model generates face IDs: starting from the \texttt{<SOS>} token, predicting the next face ID sequentially, and stopping at the \texttt{<EOS>} token, which is an autoregressive way. The number of face IDs is determined by the network automatically. In this work, the \texttt{<SOS>} token is defined as the ID of the face selected by the designer. And the \texttt{<EOS>} token is denoted by the value 0. 

During the training phase, face IDs are efficiently generated in a one-pass process, as detailed in Section~\ref{sec:training}. In contrast, during the inference phase, face IDs are generated step-by-step based on the existing (initially specified by the designer) face IDs. In the first time step, the transformer decoder inputs the \texttt{<SOS>} token corresponding face embeddings and then predicts the next face ID token using its output embeddings followed by an attention layer~\cite{vinyals2015pointer}. The attention layer in Fig.~\ref{fig:Network} takes all candidate face embeddings and decoder output embeddings as input, and computes a probability distribution across all candidate face IDs. This distribution identifies the most relevant face ID for the current time step.

Specifically, the attention outputs a probability $P(y_\tau = j 
 |  y_{<\tau},\mathbf{E}_{\mathrm{fusion}} )$ for each position \textit{j} in all face embeddings. The probability distribution is defined as: 
\begin{equation}\label{eq:attention}
P(y_\tau|y_{<\tau},\mathbf{E}_ {\mathrm{fusion}} )=\mathrm{softmax}\left( v^T \tanh( W _1\mathbf{E}_ {\mathrm{fusion}} +W_2\mathbf{h}_\tau)\right),
\end{equation}
where $W_1$,$W_2$, $v$ are learnable weights, $\mathbf{h}_\tau$ is the hidden state output by the decoder at the time step $\tau$.

At each step $\tau$, we select the position with the highest probability as the output: 
\begin{equation}
    \hat{y}_\tau=\arg\max_jP(y_\tau=\tau | y_{<\tau},\mathbf{E}_{ \mathrm{fusion}} ).
\end{equation}
After each prediction, the newly generated face ID corresponding embedding is appended to form the new decoder input for the subsequent step. Throughout decoding, an updated binary mask is used which ensures that previously predicted IDs are excluded from future selections, thereby enforcing uniqueness and coherence in the generated sequence. Thus, we can obtain generated feature label sequences \{$  \hat{y}_1,\hat{y}_2,\hat{y}_3,\ldots$\}. 

\subsection{Training}
\label{sec:training}

To accelerate the convergence of the model and reduce the error accumulation during the training process, we use the teacher forcing~\cite{williams1989learning} strategy, which directly uses the correct labels as transformer decoder input for the next time step. 

Before that, we need to build the ground-truth labels that represent the set of faces associated with each feature. The labels $ {Y} $ for a specific feature are defined as a set of face IDs representing all faces belonging to that feature. The labels can be represented as an unordered sequence of face IDs: 
\begin{equation}
     {Y} =[\texttt{<SOS>},y_1,y_2,\ldots,y_M,\texttt{<EOS>},\ldots, 0],\quad y_j\in\{1,2,\ldots,N\},
\end{equation}
where $M$ is the number of faces composing the feature, and $N$ is the total number of faces in the B-rep model. Here, the \texttt{<SOS>} token is assigned a random face ID from the set of labels Y. The length of labels ${Y}$ is the maximum number of faces that make up a feature in all input data. Each face in a specific feature shares one set of labels.
 
Specifically, at the time step $\tau$, the decoder input is the concatenation of embeddings of the first $\tau-1$ ground-truth labels: 
\begin{equation}
    \mathbf{H}_{1:\tau}=\mathbf{E}_\mathrm{fusion}[\texttt{<SOS>}, y_1,y_2,\ldots, y_{\tau-1} ].
\end{equation}
The model aims to predict the next ground-truth token $ y_{\tau}$, i.e., $P(\hat{y}_\tau=y_\tau\mid\mathbf{H}_{1:\tau},\mathbf{E}_\mathrm{fusion})$. The model parameters are optimized by minimizing the modified Binary Cross-Entropy (BCE)~\cite{ruby2020binary}  loss between the predicted results and the ground-truth labels.
 
For a batch of \textit{B} samples, each with \textit{L} fixed-length face labels, the standard BCE loss $\mathcal{L}_{\mathrm{BCE}}$ is computed as: 
\begin{equation}
    \mathcal{L}_{\mathrm{BCE}}=-\frac{1}{B\cdot L}\sum_{i=1}^{B}\sum_{j=1}^{L}\left[y_{ij}\log\hat{y}_{ij}+(1-y_{ij})\log(1-\hat{y}_{ij})\right],
\end{equation}
where ${y}_{ij} $ is the ground-truth label, and $\hat{y}_{ij} $ is the predicted probability for the \textit{j}-th face in the \textit{i}-th sample. 

In our work, the BCE loss function is specially designed to adapt to our task. Only the generated sequence before the \texttt{<EOS>} makes sense, so the loss function is applied to the result of the variable valid sequence lengths. We enhance the loss function with two key modifications:

\textbf{Masked Mean Loss Calculation.} A binary mask $\mathbf{M}\in\{0,1\}^{B\times L}$ is applied to exclude invalid positions (for positions beyond the actual length, the values are set to zero):  
\begin{equation}\label{eq:loss1}
    \mathcal{L}=-\frac{1}{\sum_{i=1}^BM_i}\sum_{i=1}^B\sum_{j=1}^L\mathbf{M}_{ij}\left[y_{ij}\log\hat{y}_{ij}+(1-y_{ij})\log(1-\hat{y}_{ij})\right],
\end{equation}
where $M _\textit{i}$ is the valid length of the \textit{i}-th sample. This ensures the loss is computed only over meaningful positions. 

\textbf{Weighted \texttt{<EOS>} Learning.} The \texttt{<EOS>} in the generated sequence is crucial as it determines the number of faces. We enhance the model's focus on \texttt{<EOS>} by higher weights, which greatly facilitates the model's accurate prediction of the number of faces in the final output. We assign a larger weight $\alpha$ to the \texttt{<EOS>} position. 
\begin{equation}\label{eq:loss2}
 \mathcal{L}_{\mathrm{final}}=\omega_{j} \mathcal{L}
\end{equation}
where the weight $\omega_{j}$ is defined as:
\begin{equation}
\omega_j = 
\begin{cases}
\alpha & \text{if position } j \text{ is the \texttt{<EOS>} token}, \\
\text{1} & \text{otherwise}.
\end{cases}
\end{equation}

This mechanism ensures that the model receives stronger feedback when correctly predicting the \texttt{<EOS>}.

\section{Results} \label{sec:results}

The evaluation of our system is conducted across two principal aspects. First, we assess the efficacy of each core module, with particular emphasis on the LLMs module (Fig.~\ref{fig:prompt_examples}), which has been enhanced through CAD-specific prompt engineering, and the performance of the feature generation module, which is evaluated on two datasets and compared with representative  method (Tab.~\ref{tab:compare}). Second, the overall system performance is demonstrated through a series of comprehensive operational examples (Fig.~\ref{fig:overallresult}).

\subsection{Setup}

\begin{figure}[t]
    \centering
    \includegraphics[width=1\linewidth]{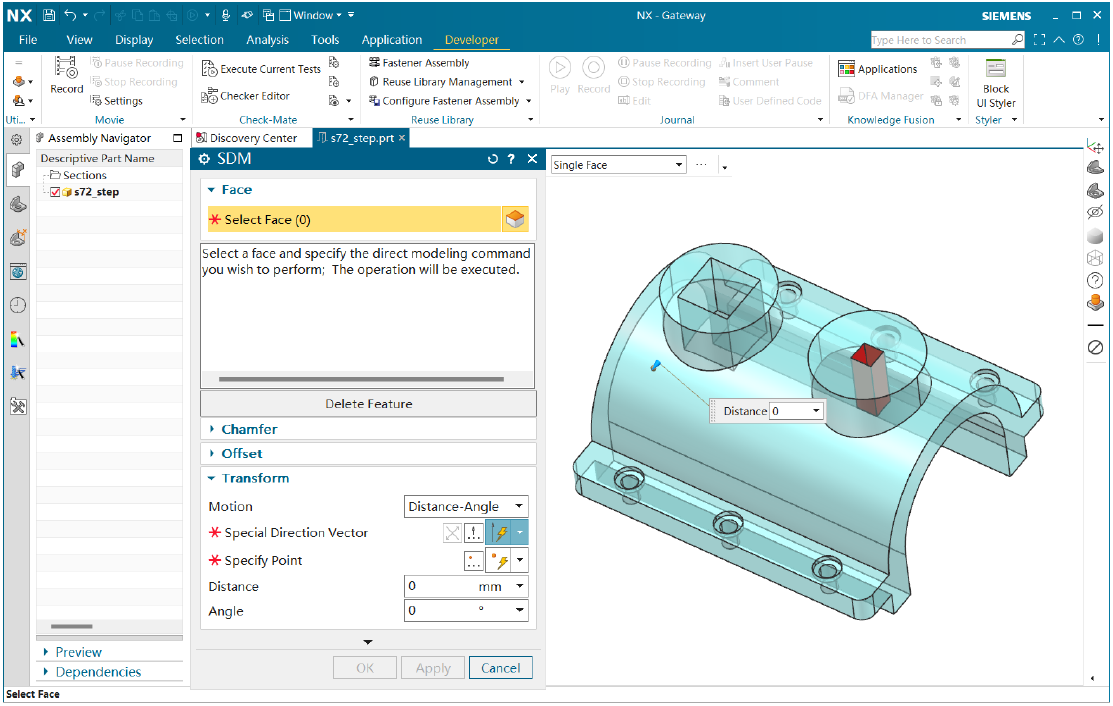}
    \caption{Customized SDM plugin development on top of Siemens NX.}
    \label{fig:NX}
\end{figure}

\textbf{Overall System Environment.} We have validated our method through customized plugin development on the commercial CAD software Siemens NX (Fig.~\ref{fig:NX}) version 2312. This development enables the software to execute speech commands from designers and offers functionality for fine-tuning operations. The Siemens NX runs on a Windows system equipped with an Intel Core i5-9th Gen CPU and NVIDIA GTX1650 GPU. Our feature generation module is trained and runs on a Linux system equipped with an Intel Core i9-12th Gen CPU and an NVIDIA RTX 4090 GPU. Similarly, the Llama and Whisper models are deployed on the same hardware. The two devices communicate with each other using socket-based communication.

\textbf{Datasets.} To verify the effectiveness of our feature generation module, we performed an evaluation on two datasets: (1) MFCAD dataset~\cite{cao2020graph} is a synthetic segmentation dataset of 15,488 3D shapes with 16 machining features. (2) MFCAD++ datasets~\cite{COLLIGAN2022103226}, which consists of 59,655 synthetic CAD models covering both planar and non-planar geometries. 

\textbf{Evaluation Metrics.} 
For the CAD-specific prompt engineering module, the output accuracy is utilized to evaluate the correctness of the model's generated outputs. This metric evaluates whether the structured outputs (e.g., operation types, target features, and parameters) fully align with the design's intent, providing a straightforward assessment of the module's performance.

For our feature generation module, the sequence generation task, we adopt the following two metrics to compare the differences between the generated results with ground-truth labels.

\begin{itemize}
    \item Intersection over Union (IoU)~\cite{zheng2020distance} measures the overlap between the generated face labels $\hat{Y}$ and the ground-truth face labels $Y$:
\begin{equation}
    \mathrm{IoU}=\frac{|\hat{Y}\cap Y|}{|\hat{Y}\cup Y|}
\end{equation}
    \item  Exact Match Rate (EM) is the proportion of generated sequences that are perfectly consistent with the ground-truth sequence:
\begin{equation}
    \mathrm{EM}=\frac1N\sum_{i=1}^N\mathbb{I}(\hat{Y}=Y),
\end{equation}
where $\mathbb{I}(\cdot)$ is the indicator function (1 for an exact match, 0 otherwise).

\end{itemize}

\subsection{Examples of CAD-specific prompt engineering}

 \begin{table}[h]
    \centering
    \caption{Accuracy of LLMs output with simple and CAD-specific prompts.}
    \begin{tabular}{llll}
        \toprule
        Methods& Simple& Complex &Total\\
        \midrule
        Simple prompts& 60\%& 0\% &15\%\\
          CAD-specific prompts& 100\%& 96.67\% &97.5\%\\
        \bottomrule
    \end{tabular}
    \vspace{0.2em}
    \label{tab:prompt_rates}  
\end{table}

\begin{figure*}[h!]
    \centering
    \includegraphics[width=0.95\linewidth]{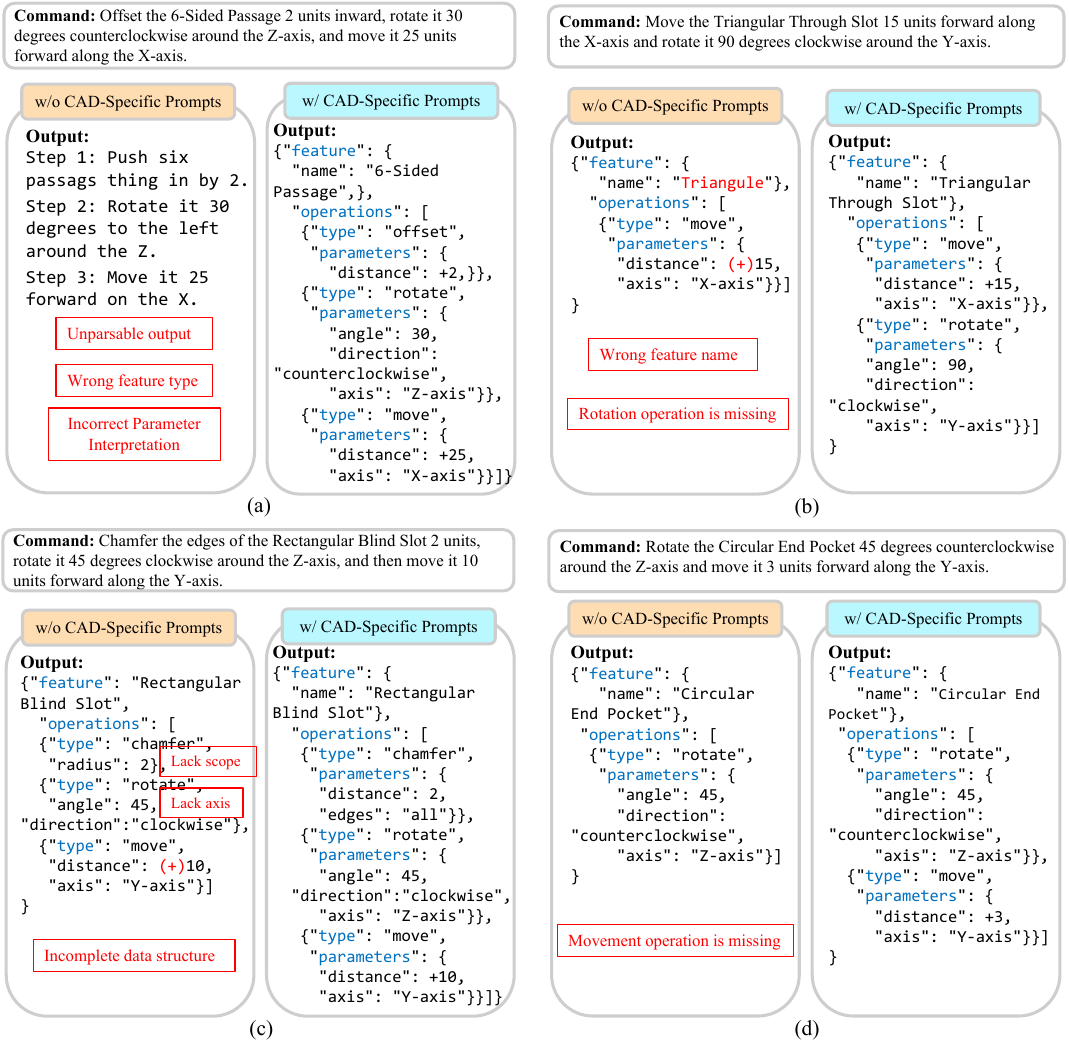}
     \caption{Comparative analysis of LLMs outputs using simple prompts versus CAD-specific prompts across four case studies.}
  \label{fig:prompt_examples}
\end{figure*}

\begin{figure*}[hpb]
    \centering
    
    \includegraphics[width=0.93\linewidth]{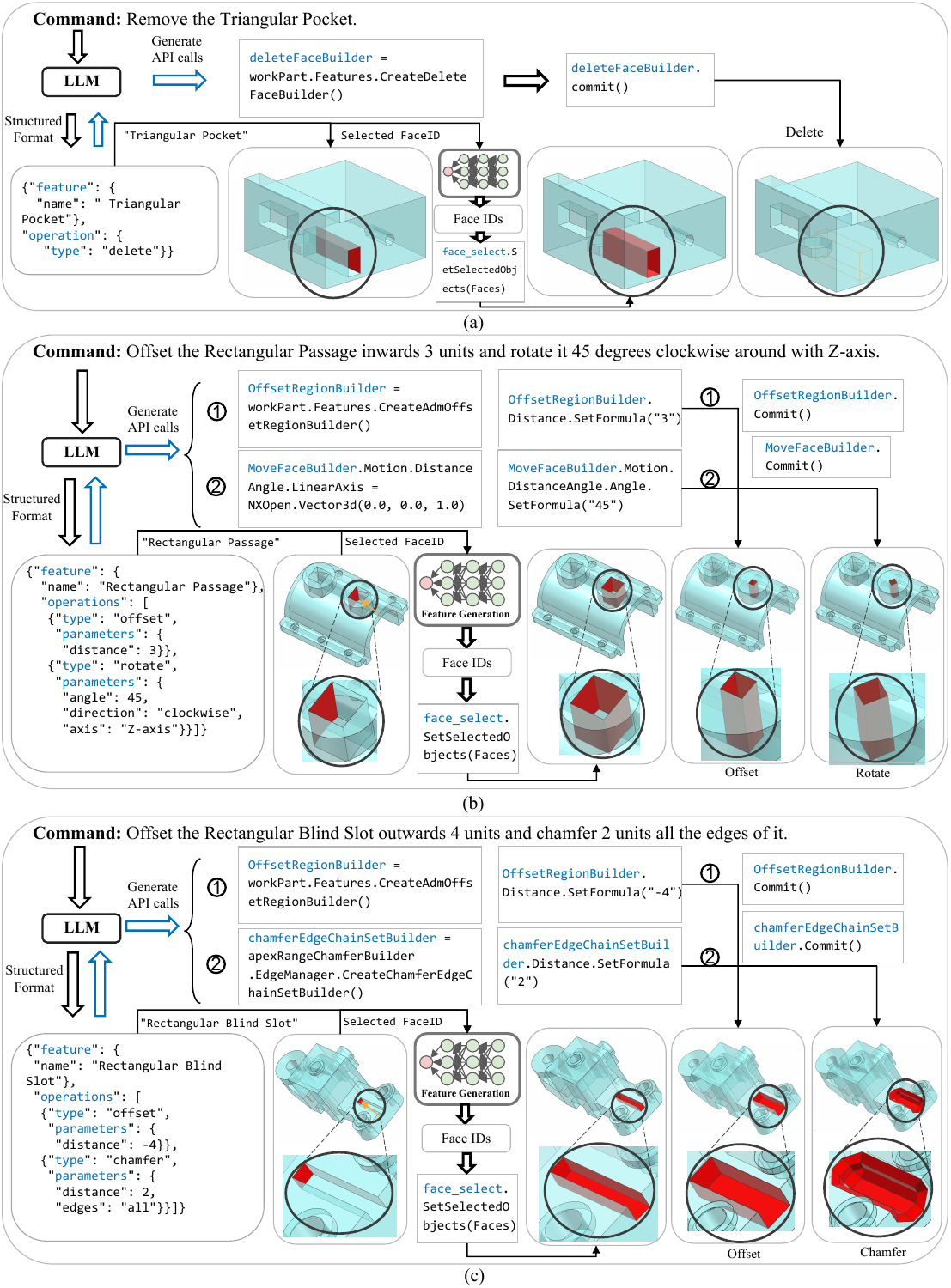}
   \caption{Illustration of the SDM system's workflow across three examples. Each example demonstrates the complete process, from LLM interpretation of commands to the generation of API snippets and the subsequent model transformations.}
    \label{fig:overallresult}
   
\end{figure*}

We constructed a comprehensive set of 40 designer descriptions, encompassing a diverse range of operational scenarios, which includes 10 simple cases and 30 complex cases, each associated with a unique correct response. The accuracy rates for these cases are detailed in Tab.~\ref{tab:prompt_rates}. Four representative case studies 1-4, illustrating the comparison between CAD-specific prompts and simple prompts, are shown in Fig.~\ref{fig:prompt_examples}a-d. Case studies 1,3 (Fig.~\ref{fig:prompt_examples}a,c) have three operation types and corresponding parameters. Case studies 2,4 (Fig.~\ref{fig:prompt_examples}b,d) have two.

\subsection{Feature generation module performance}
 \begin{table}[h!]
    \centering
    \caption{Comparative performance on feature generation tasks for the MFCAD and MFCAD++ datasets.}\begin{tabular}{llccc}
        \toprule
        \text{Dataset} & \text{Network}   & \text{IoU (\%)} & \text{ME (\%)} & \text{Params (M)} \\
        \midrule
        \multirow{3}{*}{\textit{MFCAD}}
          & Ours& 99.54& 99.20& 9.8\\
          & UV-Net & 98.59& 97.46& 6.5\\
        \addlinespace[0.1em]\multirow{3}{*}{\textit{MFCAD++}}& Ours& 98.50& 96.73& 9.8\\
          & UV-Net & 96.78& 93.27& 6.5\\
        \bottomrule
    \end{tabular}
    \vspace{0.2em}
    
    \label{tab:compare}  
\end{table}
To adapt these datasets to our method, we employ the OpenCascade open-source library, a widely recognized tool in geometric processing, for preprocessing. Specifically, each B-rep face is converted into a mesh, and non-manifold models are filtered out during this process. Following preprocessing, the MFCAD++ dataset comprises 57,992 models, whereas the MFCAD dataset remains unchanged. 

The comparative results are presented in Tab.~\ref{tab:compare}. As previously mentioned, existing methods primarily rely on discrete approaches, utilizing GNNs and CNNs as encoders for feature recognition, with UV-Net being the most representative method. In contrast, our proposed approach introduces a conditional generative framework. To rigorously validate this design, we conducted module-level comparative experiments. Specifically, we integrated our conditional generative module into UV-Net and evaluated both variants on the MFCAD and MFCAD++ datasets.



\subsection{Examples of the overall system}

Several real-world design tasks have been implemented to demonstrate the practical applicability of the methods proposed in this paper. As illustrated in Fig.~\ref{fig:overallresult}a-c, intuitive workflow case studies have been provided, enabling designers to select faces using straightforward click-and-speech commands. Case study 1 (Fig.~\ref{fig:overallresult}a) includes a single simple operation (``Delete") without parameters. Case studies 2-3 (Fig.~\ref{fig:overallresult}b-c) each encompass two operations with specified parameters, such as direction, distance, and angle. The structured outputs generated by the LLMs after processing these commands and the secondary development code snippets obtained after re-inputting the parameters are also presented.


\subsection{Discussion and limitations}

In our LLMs module, as shown in Tab.~\ref{tab:prompt_rates}, CAD-specific prompt engineering achieves an accuracy rate of 97.5\% with only 1 error, compared to 15\% accuracy (6 correct responses) using simple prompts. Fig.~\ref{fig:prompt_examples} illustrates that multi-step commands (e.g., ``Rotate... and move...") are often incompletely or incorrectly parsed by simple prompts, resulting in inconsistent and unreliable outputs. In contrast, LLMs with CAD-specific prompts consistently produce structured, accurate outputs, crucial for robust CAD command parsing and automated system integration.

From the performance metrics presented in Tab.~\ref{tab:compare}, our feature generation module achieves high accuracy rates on both the MFCAD and MFCAD++ datasets, with IoU values of 99.54\% and 98.50\%, respectively. These results are satisfactory for practical designer use. Compared to UV-Net, which achieves IoU values of 98.59\% on MFCAD and 96.78\% on MFCAD++, our method consistently outperforms various metrics and datasets. 

Even though the overall system has demonstrated its effectiveness through specific cases in Fig.~\ref{fig:overallresult}a-c, there is a notable limitation. In practical testing, we found that the overall efficiency of the workflow, which involves communication between the direct modeling software and the AI, is constrained by network conditions. It would be more effective to embed a lightweight version of the model directly into the software.

\section{Conclusion} \label{sec:conclusion}

This paper introduces the SDM system, an innovative exploration of integrating CAD direct modeling with LLMs and generative AI. The system bridges high-level design intent with low-level geometric modifications in CAD systems through the following two aspects: 
first, by employing a fine-tuned LLM with CAD-specific prompts, SDM accurately interprets CAD commands, thus facilitating the natural language expression of design intent by designers; second, SDM utilizes a novel conditional, context-sensitive feature recognition method that leverages generative AI to generate features IDs based on design intent dynamically. The effectiveness of SDM has been validated through a series of mechanical design examples of increasing shape complexity. 

Although the proposed method represents a significant advancement in direct modeling, there are still several limitations. One of the primary challenges is that LLMs currently struggle with comprehending instructions involving spatial relationships such as ``Move the Circular Hole feature to the upper left corner of the main body", primarily because the LLMs lack direct access to the specific spatial state of the model. Integrating real-time image data from the interaction interface as an extra modality could potentially address this limitation, and it is among our future research studies. Another limitation to note is that while prompt engineering has proven adequate for the majority of scenarios, it is not capable of inputting all the knowledge of a specific domain at once. The Retrieval-Augmented Generation (RAG)~\cite{lewis2020retrieval} method presents a potential solution to this issue; however, further development is necessary to refine its capabilities fully. Additionally, the current SDM system focuses solely on incorporating design intent into direct modeling, overlooking analysis intent (such as parametric optimization~\cite{tang2023decision,li2023xvoxel}) and manufacturing requirements (e.g., tool path generation~\cite{zou2014iso,zou2021length,wang2023computing}). Expanding its scope to include these aspects is a key interest of our research.

\section*{Acknowledgements}
This work has been funded by the NSF of China (No. 62102355) and the ``Pioneer” and ``Leading Goose” R\&D Program of Zhejiang Province, China (No. 2024C01103).

 \bibliographystyle{elsarticle-num} 
 \bibliography{cas-refs}

\begin{thebibliography}{10}
\expandafter\ifx\csname url\endcsname\relax
  \def\url#1{\texttt{#1}}\fi
\expandafter\ifx\csname urlprefix\endcsname\relax\def\urlprefix{URL }\fi
\expandafter\ifx\csname href\endcsname\relax
  \def\href#1#2{#2} \def\path#1{#1}\fi

\bibitem{li2020computer}
W.~Li, G.~Mac, N.~G. Tsoutsos, N.~Gupta, R.~Karri, Computer aided design ({CAD}) model search and retrieval using frequency domain file conversion, Additive Manufacturing 36 (2020) 101554.

\bibitem{zou2024meta}
Q.~Zou, Y.~Gao, G.~Luo, S.~Chen, Meta-meshing and triangulating lattice structures at a large scale, Computer-Aided Design 174 (2024) 103732.

\bibitem{lee1999principles}
K.~Lee, Principles of CAD/CAM/CAE systems, Addison-Wesley Longman Publishing Co., Inc., 1999.

\bibitem{zhao2024tpms2step}
Y.~Zhao, Q.~Zou, G.~Luo, J.~Wu, S.~Chen, D.~Gao, M.~Xuan, F.~Wang, Tpms2step: error-controlled and c2 continuity-preserving translation of tpms models to step files based on constrained-pia, Computer-Aided Design 173 (2024) 103726.

\bibitem{shapiro2002solid}
V.~Shapiro, Solid modeling., Handbook of computer aided geometric design 20 (2002) 473--518.

\bibitem{shah1995parametric}
J.~J. Shah, M.~M{\"a}ntyl{\"a}, Parametric and feature-based CAD/CAM: concepts, techniques, and applications, John Wiley \& Sons, 1995.

\bibitem{zou2023variational}
Q.~Zou, H.-Y. Feng, S.~Gao, Variational direct modeling: A framework towards integration of parametric modeling and direct modeling in {CAD}, Computer-Aided Design 157 (2023) 103465.

\bibitem{zou2019push}
Q.~Zou, H.-Y. Feng, Push-pull direct modeling of solid {CAD} models, Advances in Engineering Software 127 (2019) 59--69.

\bibitem{ault2016direct}
H.~K. Ault, A.~Phillips, Direct modeling: easy changes in {CAD}? (2016).

\bibitem{li2023xvoxel}
M.~Li, C.~Lin, W.~Chen, Y.~Liu, S.~Gao, Q.~Zou, Xvoxel-based parametric design optimization of feature models, Computer-Aided Design 160 (2023) 103528.

\bibitem{zou2013iso}
Q.~Zou, J.~Zhao, Iso-parametric tool-path planning for point clouds, Computer-Aided Design 45~(11) (2013) 1459--1468.

\bibitem{zhang2025diffusion}
A.~Zhang, W.~Jia, Q.~Zou, Y.~Feng, X.~Wei, Y.~Zhang, Diffusion-cad: Controllable diffusion model for generating computer-aided design models, IEEE Transactions on Visualization and Computer Graphics (2025).

\bibitem{zou2025splinegen}
Q.~Zou, L.~Zhu, J.~Wu, Z.~Yang, Splinegen: Approximating unorganized points through generative ai, Computer-Aided Design 178 (2025) 103809.

\bibitem{bidarra2000semantic}
R.~Bidarra, W.~F. Bronsvoort, Semantic feature modelling, Computer-Aided Design 32~(3) (2000) 201--225.

\bibitem{jayaraman2021uv}
P.~K. Jayaraman, A.~Sanghi, J.~G. Lambourne, K.~D. Willis, T.~Davies, H.~Shayani, N.~Morris, {UV}-net: Learning from boundary representations, in: Proceedings of the IEEE/CVF conference on computer vision and pattern recognition, 2021, pp. 11703--11712.

\bibitem{COLLIGAN2022103226}
A.~R. Colligan, T.~T. Robinson, D.~C. Nolan, Y.~Hua, W.~Cao, Hierarchical {CADNet}: Learning from b-reps for machining feature recognition, Computer-Aided Design 147 (2022) 103226.

\bibitem{zhang2024brepmfr}
S.~Zhang, Z.~Guan, H.~Jiang, X.~Wang, P.~Tan, Brepmfr: Enhancing machining feature recognition in b-rep models through deep learning and domain adaptation, Computer Aided Geometric Design 111 (2024) 102318.

\bibitem{wu2024aagnet}
H.~Wu, R.~Lei, Y.~Peng, L.~Gao, Aagnet: A graph neural network towards multi-task machining feature recognition, Robotics and Computer-Integrated Manufacturing 86 (2024) 102661.

\bibitem{lambourne2021brepnet}
J.~G. Lambourne, K.~D. Willis, P.~K. Jayaraman, A.~Sanghi, P.~Meltzer, H.~Shayani, Brepnet: A topological message passing system for solid models, in: Proceedings of the IEEE/CVF conference on computer vision and pattern recognition, 2021, pp. 12773--12782.

\bibitem{grayer1980alternative}
A.~R. Grayer, Alternative approaches in geometric modelling, Computer-Aided Design 12~(4) (1980) 189--192.

\bibitem{stroud2006boundary}
I.~Stroud, Boundary representation modelling techniques, Springer Science \& Business Media, 2006.

\bibitem{fahlbusch1995hp}
K.-P. Fahlbusch, T.~D. Roser, Hp pe/soliddesigner: dynamic modeling for three-dimensional computer-aided design, Hewlett-Packard Journal 46~(5) (1995) 6--13.

\bibitem{rossignac1990issues}
J.~R. Rossignac, Issues on feature-based editing and interrogation of solid models, Computers \& Graphics 14~(2) (1990) 149--172.

\bibitem{zou2022robust}
Q.~Zou, H.-Y. Feng, A robust direct modeling method for quadric b-rep models based on geometry--topology inconsistency tracking, Engineering with Computers 38~(4) (2022) 3815--3830.

\bibitem{lipp2014pushpull}
M.~Lipp, P.~Wonka, P.~M{\"u}ller, Pushpull++, ACM Transactions on Graphics (TOG) 33~(4) (2014) 1--9.

\bibitem{zou2020decision}
Q.~Zou, H.-Y. Feng, A decision-support method for information inconsistency resolution in direct modeling of {CAD} models, Advanced Engineering Informatics 44 (2020) 101087.

\bibitem{pan2022automatic}
W.~Pan, C.~Chen, Y.~Yang, S.~Gao, Y.~Wang, S.~Wang, Automatic shape adaptation scheme planning for {CAD} models in direct modeling, Computer-Aided Design 153 (2022) 103405.

\bibitem{zou2019variational}
Q.~Zou, H.-Y. Feng, Variational b-rep model analysis for direct modeling using geometric perturbation, Journal of Computational Design and Engineering 6~(4) (2019) 606--616.

\bibitem{zou2024intelligent}
Q.~Zou, Y.~Wu, Z.~Liu, W.~Xu, S.~Gao, Intelligent {CAD 2.0}, Visual Informatics (2024).

\bibitem{lee2023brepgat}
J.~Lee, C.~Yeo, S.-U. Cheon, J.~H. Park, D.~Mun, Brepgat: Graph neural network to segment machining feature faces in a b-rep model, Journal of Computational Design and Engineering 10~(6) (2023) 2384--2400.

\bibitem{Lou2023brepbert}
Y.~Lou, X.~Li, H.~Chen, X.~Zhou, Brep-bert: Pre-training boundary representation bert with sub-graph node contrastive learning, in: Proceedings of the 32nd ACM International Conference on Information and Knowledge Management, Association for Computing Machinery, 2023, p. 1657–1666.

\bibitem{liu2023meshdiffusion}
Z.~Liu, Y.~Feng, M.~J. Black, D.~Nowrouzezahrai, L.~Paull, W.~Liu, Meshdiffusion: Score-based generative 3d mesh modeling, arXiv preprint arXiv:2303.08133 (2023).

\bibitem{chen2025meshxl}
S.~Chen, X.~Chen, A.~Pang, X.~Zeng, W.~Cheng, Y.~Fu, F.~Yin, B.~Wang, J.~Yu, G.~Yu, et~al., Meshxl: Neural coordinate field for generative 3d foundation models, Advances in Neural Information Processing Systems 37 (2025) 97141--97166.

\bibitem{luo2021diffusion}
S.~Luo, W.~Hu, Diffusion probabilistic models for 3d point cloud generation, in: Proceedings of the IEEE/CVF conference on computer vision and pattern recognition, 2021, pp. 2837--2845.

\bibitem{shu20193d}
D.~W. Shu, S.~W. Park, J.~Kwon, 3d point cloud generative adversarial network based on tree structured graph convolutions, in: Proceedings of the IEEE/CVF international conference on computer vision, 2019, pp. 3859--3868.

\bibitem{xu2024brepgen}
X.~Xu, J.~Lambourne, P.~Jayaraman, Z.~Wang, K.~Willis, Y.~Furukawa, Brepgen: A b-rep generative diffusion model with structured latent geometry, ACM Transactions on Graphics (TOG) 43~(4) (2024) 1--14.

\bibitem{guo2022complexgen}
H.~Guo, S.~Liu, H.~Pan, Y.~Liu, X.~Tong, B.~Guo, Complexgen: {CAD} reconstruction by b-rep chain complex generation, ACM Transactions on Graphics (TOG) 41~(4) (2022) 1--18.

\bibitem{jayaraman2023solidgen}
P.~K. Jayaraman, J.~G. Lambourne, N.~Desai, K.~Willis, A.~Sanghi, N.~J.~W. Morris, Solidgen: An autoregressive model for direct b-rep synthesis, Transactions on Machine Learning Research (2023).

\bibitem{wu2021deepcad}
R.~Wu, C.~Xiao, C.~Zheng, Deepcad: A deep generative network for computer-aided design models, in: Proceedings of the IEEE/CVF International Conference on Computer Vision, 2021, pp. 6772--6782.

\bibitem{nx}
{NX}, \url{https://plm.sw.siemens.com/en-US/nx/}, accessed: 2025-02-25.

\bibitem{ansys}
{ANSYS-SpaceClaim}, \url{https://ansys.com/products/3d-design/ansys-spaceclaim}, accessed: 2025-01-28.

\bibitem{ptccreo}
{PTC Creo}, \url{https://www.ptc.com/en/products/creo}, accessed: 2025-01-13.

\bibitem{autodesk}
{Autodesk}, \url{https://www.autodesk.com/products/inventor/}, accessed: 2025-01-07.

\bibitem{dubey2024llama}
A.~Dubey, A.~Jauhri, A.~Pandey, A.~Kadian, A.~Al-Dahle, A.~Letman, A.~Mathur, A.~Schelten, A.~Yang, A.~Fan, et~al., The llama 3 herd of models, arXiv preprint arXiv:2407.21783 (2024).

\bibitem{radford2023robust}
A.~Radford, J.~W. Kim, T.~Xu, G.~Brockman, C.~McLeavey, I.~Sutskever, Robust speech recognition via large-scale weak supervision, in: International conference on machine learning, PMLR, 2023, pp. 28492--28518.

\bibitem{vaswani2017attention}
A.~Vaswani, Attention is all you need, Advances in Neural Information Processing Systems (2017).

\bibitem{jiang2024survey}
J.~Jiang, F.~Wang, J.~Shen, S.~Kim, S.~Kim, A survey on large language models for code generation, arXiv preprint arXiv:2406.00515 (2024).

\bibitem{raiaan2024review}
M.~A.~K. Raiaan, M.~S.~H. Mukta, K.~Fatema, N.~M. Fahad, S.~Sakib, M.~M.~J. Mim, J.~Ahmad, M.~E. Ali, S.~Azam, A review on large language models: Architectures, applications, taxonomies, open issues and challenges, IEEE Access (2024).

\bibitem{schick2020exploiting}
T.~Schick, H.~Sch{\"u}tze, Exploiting cloze questions for few shot text classification and natural language inference, arXiv preprint arXiv:2001.07676 (2020).

\bibitem{brown2020language}
T.~Brown, B.~Mann, N.~Ryder, M.~Subbiah, J.~D. Kaplan, P.~Dhariwal, A.~Neelakantan, P.~Shyam, G.~Sastry, A.~Askell, et~al., Language models are few-shot learners, Advances in neural information processing systems 33 (2020) 1877--1901.

\bibitem{wei2022chain}
J.~Wei, X.~Wang, D.~Schuurmans, M.~Bosma, F.~Xia, E.~Chi, Q.~V. Le, D.~Zhou, et~al., Chain-of-thought prompting elicits reasoning in large language models, Advances in neural information processing systems 35 (2022) 24824--24837.

\bibitem{dosovitskiy2020image}
A.~Dosovitskiy, L.~Beyer, A.~Kolesnikov, D.~Weissenborn, X.~Zhai, T.~Unterthiner, M.~Dehghani, M.~Minderer, G.~Heigold, S.~Gelly, et~al., An image is worth 16x16 words: Transformers for image recognition at scale, arXiv preprint arXiv:2010.11929 (2020).

\bibitem{feng2019meshnet}
Y.~Feng, Y.~Feng, H.~You, X.~Zhao, Y.~Gao, Meshnet: Mesh neural network for 3d shape representation, in: Proceedings of the AAAI conference on artificial intelligence, Vol.~33, 2019, pp. 8279--8286.

\bibitem{vinyals2015pointer}
O.~Vinyals, M.~Fortunato, N.~Jaitly, Pointer networks, Advances in neural information processing systems 28 (2015).

\bibitem{williams1989learning}
R.~J. Williams, D.~Zipser, A learning algorithm for continually running fully recurrent neural networks, Neural computation 1~(2) (1989) 270--280.

\bibitem{ruby2020binary}
U.~Ruby, V.~Yendapalli, Binary cross entropy with deep learning technique for image classification, Int. J. Adv. Trends Comput. Sci. Eng 9~(10) (2020).

\bibitem{cao2020graph}
W.~Cao, T.~Robinson, Y.~Hua, F.~Boussuge, A.~R. Colligan, W.~Pan, Graph representation of 3d {CAD} models for machining feature recognition with deep learning, in: International design engineering technical conferences and computers and information in engineering conference, Vol. 84003, American Society of Mechanical Engineers, 2020, p. V11AT11A003.

\bibitem{zheng2020distance}
Z.~Zheng, P.~Wang, W.~Liu, J.~Li, R.~Ye, D.~Ren, Distance-iou loss: Faster and better learning for bounding box regression, in: Proceedings of the AAAI conference on artificial intelligence, Vol.~34, 2020, pp. 12993--13000.

\bibitem{lewis2020retrieval}
P.~Lewis, E.~Perez, A.~Piktus, F.~Petroni, V.~Karpukhin, N.~Goyal, H.~K{\"u}ttler, M.~Lewis, W.-t. Yih, T.~Rockt{\"a}schel, et~al., Retrieval-augmented generation for knowledge-intensive nlp tasks, Advances in neural information processing systems 33 (2020) 9459--9474.

\bibitem{tang2023decision}
Z.~Tang, Q.~Zou, S.~Gao, A decision-support method for multi-parameter editing of parametric cad models, Advanced Engineering Informatics 56 (2023) 101997.

\bibitem{zou2014iso}
Q.~Zou, J.~Zhang, B.~Deng, J.~Zhao, Iso-level tool path planning for free-form surfaces, Computer-Aided Design 53 (2014) 117--125.

\bibitem{zou2021length}
Q.~Zou, Length-optimal tool path planning for freeform surfaces with preferred feed directions based on poisson formulation, Computer-Aided Design 139 (2021) 103072.

\bibitem{wang2023computing}
Z.~Wang, S.~Liu, L.~Liu, Q.~Zou, X.-M. Fu, Computing smooth preferred feed direction fields with high material removal rates for efficient cnc tool paths, Computer-Aided Design 164 (2023) 103591.

\end{thebibliography}





\end{document}